\documentclass[superscriptaddress,showpacs,twocolumn,aps,pre,10pt]{revtex4-1}
\usepackage{natbib}
\usepackage{graphicx,dcolumn,bm,bbm,amsmath,amssymb,color}

\renewcommand{\vec}[1]{\boldsymbol{#1}}%
\newcommand{\bea}{\begin{eqnarray}}
\newcommand{\eea}{\end{eqnarray}}

\newcommand{\bbr}{ {\bf r} }
\newcommand{\bhu}{ \hat{\bf u} }

\newcommand{\kbt}{k_{\rm B}T}

\begin{document}
\title{\Large{How does a flexible chain of active particles swell?}\vspace{0.3cm}}

\author{Andreas Kaiser}
\email{kaiser@thphy.uni-duesseldorf.de}
\affiliation{Institut f\"ur Theoretische Physik II: Weiche Materie,
Heinrich-Heine-Universit\"at D\"{u}sseldorf,
Universit{\"a}tsstra{\ss}e 1, D-40225 D\"{u}sseldorf, Germany}
\author{Sonja Babel}
\affiliation{Institut f\"ur Theoretische Physik II: Weiche Materie,
Heinrich-Heine-Universit\"at D\"{u}sseldorf,
Universit{\"a}tsstra{\ss}e 1, D-40225 D\"{u}sseldorf, Germany}
\author{Borge ten Hagen}
\affiliation{Institut f\"ur Theoretische Physik II: Weiche Materie,
Heinrich-Heine-Universit\"at D\"{u}sseldorf,
Universit{\"a}tsstra{\ss}e 1, D-40225 D\"{u}sseldorf, Germany}
\author{Christian von Ferber}
\affiliation{Institut f\"ur Theoretische Physik II: Weiche Materie,
Heinrich-Heine-Universit\"at D\"{u}sseldorf,
Universit{\"a}tsstra{\ss}e 1, D-40225 D\"{u}sseldorf, Germany}
\affiliation{Applied Mathematics Research Centre, Coventry University, Coventry CV1 5FB, UK}
\author{Hartmut L\"{o}wen}
\affiliation{Institut f\"ur Theoretische Physik II: Weiche Materie,
Heinrich-Heine-Universit\"at D\"{u}sseldorf,
Universit{\"a}tsstra{\ss}e 1, D-40225 D\"{u}sseldorf,
Germany}

\date{\today}


\begin{abstract}
We study the swelling of a flexible linear chain composed of active particles 
by analytical theory and computer simulation. Three different situations are 
considered: a free chain, a chain confined to an external
harmonic trap, and a chain dragged at one end.
First we consider an ideal chain with harmonic springs and
no excluded volume between the monomers. The Rouse model of polymers  
is generalized to the case of self-propelled monomers and solved analytically.
 The swelling, as characterized by the spatial extension of
the chain, scales with the monomer number defining
a Flory exponent $\nu$ which is  $\nu =1/2, 0, 1$ in the three different situations.
As a result, we find that activity does not change the Flory
exponent but affects the prefactor of the scaling law. This can be quantitatively
understood by mapping the system onto an equilibrium chain
with a higher effective temperature such that the chain swells under an increase of the self-propulsion strength. 
We then use computer simulations to study the effect of self-avoidance 
on active polymer swelling. In the three different situations, the
Flory exponent is now $\nu = 3/4, 1/4, 1$ and again unchanged under self-propulsion. 
However, the chain extension behaves non-monotonic in the self-propulsion strength.
\end{abstract}

\maketitle

\section{Introduction}
\label{sec:intro}

In recent years, much interdisciplinary research in soft matter physics,
fluid mechanics and biology has been devoted to understand the motion of microswimmers
in a low-Reynolds-number fluid~\cite{Romanczuk2012,Marchetti_Rev,Cates_Rev,aranson_ufn,Menzel_Rev,Gompper_Rev}.
Microswimmers can either be found as microbes such as bacteria~\cite{PoonEmulsionDrop}, viruses or algae,~\cite{GoldsteinVolvox}
or are realized artificially as self-propelled (``active'')
colloidal particles~\cite{PaxtonJACS2006,Bechinger_SM11,ErbeBaraban2008,BocquetPRL2010,Palacci_science}.
While the former
are typically self-propelled by changing their shape, e.g. using beating flexible flagella \cite{BMF_PRL14,Leptos_PRL14},
the latter are form-stable Janus particles
exposed to a self-generated  chemical or thermal gradient which brings the particle into motion \cite{KapralJCP2013,Sano_PRL2010,Bechinger}.
The combination of self-propulsion and rotational Brownian motion ultimately leads to diffusion
however with a much higher
diffusion coefficient as compared to unpropelled
 (``passive'') particles~\cite{Howse_2007,BtH2011}.

Along the route of recent research, the shape of artificial colloidal swimmers
has been made more complex by considering 
Janus-spheres~\cite{BocquetPRL2010,Bechinger,Bialke_PRL2013,Reichhardt_PRL2008,ZoettelPRL14,Golestanian_2012,Egorov_PRL14,Chen_JCP14,Ni}, 
rods~\cite{PaxtonJACS2004,2006Peruani,PNAS,KaiserIEEE,SanchezJACS2011} and particles
of arbitrary shape~\cite{Wittkowski2012,BtH_L_part,Menzel_EPL,MiniRev_D3,BtHNatComm,Wensink2014}. 
At the same time, the flexibility of the flagella was incorporated in models for microbe motion
~\cite{Netz_EPL08,2010Gompper,SpermAlvarez,SpermsFriedrich}.

Here, we consider a flexible object that is composed of
many active constituents, namely a linear chain of self-propelled particles. 
The motivation to do so is threefold:
first, in a general sense, it is necessary to study how the collective behavior
of microswimmers depends on their mutual coupling. Typically a pairwise interaction
potential is assumed between all microswimmers but what is barely understood is how
this behavior is affected by a specific strong coupling topology defining a connectivity,
e.g. along a linear chain.
Secondly, chains of active monomers are at the interface between the physics of microswimmers and polymer science
such that they establish a natural link between these two scientific disciplines.
It would indeed be challenging to generalize the broad concepts of polymer scaling theory
~\cite{deGennesbook,Doi_Edwards_book,book_by_Lothar_Schaefer}
towards nonequilibrium~\cite{Netz2012,WnklerPolyShearFlow06,Muthukumar2004,WinklerGompper2008} and the physics
of self-propelled particles.
Lastly, polymers of active particles can be realized by chaining artificial colloids, using e.g. the
lock-and-key technique~\cite{Pine_Nature_2010} or DNA~\cite{Bibette}, towards a linear chain of colloids. Recently, these linear colloidal polymers 
have been prepared and called ``polloidal chains''~\cite{polloidal_chains} or ``colloidal caterpillars''\cite{Herminghaus_SM14}.
When these colloidal beads are replaced by active colloids, the situation of a flexible chain 
considered here is in principle experimentally realizable
such that theoretical predictions can be verified on the monomer-resolved level. 
Another realization of an active chain is a shaken granulate chain which has already been
realized~\cite{ExpGranularActiveChain1,ExpGranularActiveChain2}. Here the millimetric beads can
be studied in real-space. Though the details of our modeling apply to microswimmers in a solvent,
we expect qualitative similarities between an active polymer solution and an active granular chain.

In previous work, appropriate models 
for an active semiflexible chain have been studied. Here a finite persistence length
along the chain is assumed but the chain is typically short in the sense that it does not reach
its limit of coiling. Some of the previous studies \cite{Loi_SM1,Loi_SM2}
consider only a single active bead along the polymer, others focus on the dynamics  
of an active semi-flexible chain using either simulation or field theory \cite{Jayaraman_2012,Hagan_2013,Gov_2014}.
Lastly, very recently, a one-dimensional chain of active beads (i.e. the case of perfect persistence)
has been studied in a ratchet potential \cite{EPJB}.

In this paper, we focus on the swelling behavior of a flexible chain composed of active beads 
using both analytical theory and computer simulation. In doing so we discriminate between an ideal chain 
(i.e. a chain without self-avoidance) and a self-avoiding chain. For the former case, we 
generalize the traditional Rouse model~\cite{Rouse_original,Doi_Edwards_book}
 of polymer dynamics towards
the situation of active monomers, for the latter we use computer simulations. 
In the spirit of the simplicity of the Rouse model which assumes
a linear bead-spring chain with harmonic coupling between nearest neighbors and neglected hydrodynamic interactions
we introduce an activity for the monomers but neglect any correlations in the activity of neighboring
beads. This generalized Rouse model with no explicit aligning interactions 
for active polymers is then solved analytically in two dimensions.
As a basic result, the Flory random-walk exponent $\nu =1/2$ which measures how
the typical extension of the chain scales with
the monomer number is not affected by activity but the prefactor is corresponding to a higher
effective temperature when mapped onto a corresponding passive chain. This implies that the
extension of an active chain increases with the strength of the self-propulsion (or P\'eclet number).

Self-avoidance in two dimensions yields a Flory swelling exponent $\nu =3/4$ for passive monomers.
Our computer simulations show that for long chains this exponent is not affected by self-propulsion of the monomers.
This is in line with experiments using a shaken granular chain~\cite{ExpGranularActiveChain1,ExpGranularActiveChain2}
where the Flory exponent was found to be unaffected by the activity.
However, as a function of the self-propulsion strength, there is a non-monotonicity in the chain extension
which is absent for an ideal chain. 
Interestingly, this effect was found recently
in the reverse set-up  of a passive chain in an active bath~\cite{KaiserPolymer,ValerianiPolymer}
which can be realized for granulates~\cite{ExpGranularChain}, and which shows the same Flory scaling for large monomer number.

We also solve the Rouse model of an active ideal chain either confined to an external
harmonic trap or dragged at one end. In these situations the
Flory exponent is  $\nu =0$ and $\nu=1$, respectively, and also not changed by activity.
But again the prefactor is affected by self-propulsion. Self-avoidance leads to the Flory exponents 
$\nu =1/4$ and $\nu=1$ in these two situations. Our simulations show that these exponents are 
unaffected by internal activity of the chain.

The paper is organized as follows: we describe our model and the different situations in Sec.~\ref{sec:model} and the simulation
in Sec.~\ref{sec:simu}. Results are discussed in Sec.~\ref{sec:results}, and we conclude in Sec.~\ref{sec:conc} where we
comment on how an active polymer can be realized experimentally.

\section{Rouse model for an ideal chain of active particles}
\label{sec:model}

\begin{figure}[thb]
\begin{center}
\includegraphics[width=1\columnwidth]{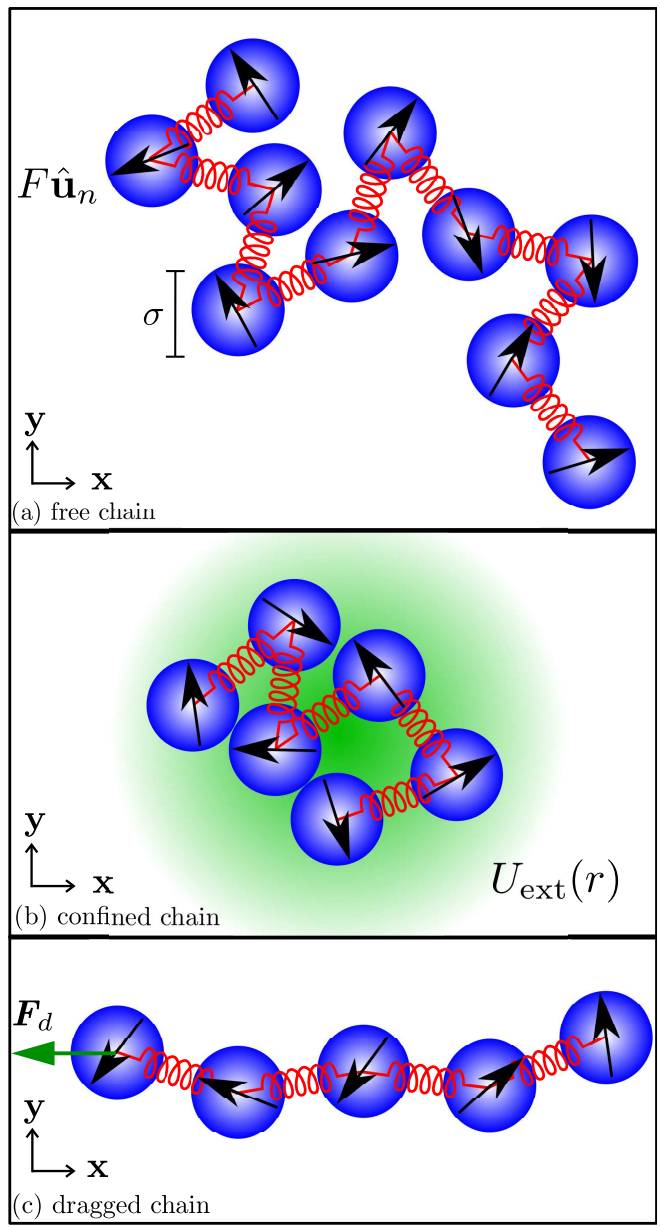}
\caption{\label{fig:Sketch} Sketch of an active polymer consisting of $N$ monomers which are self-propelled 
with an effective driving force $F$ along their orientation $\bhu_n$ for the three studied situations: (a) free chain, 
(b) chain in harmonic confinement $U_{\text{ext}}$ and (c)  chain with a dragging force $\vec{F}_d$ applied to the first monomer.}
\end{center}
\end{figure}

In our Rouse-like model we describe the polymer as a linear chain of $N$ beads. Neighbors are connected 
by harmonic springs~\footnote{Similar bead-spring 
models with a two-dimensional connectivity were studied in the context of active crystals by Huepe and coworkers~\cite{Huepe}.} 
with a spring constant $k$. We focus on the two dimensional case. At a given time $t$, the monomers are at  positions 
$\vec{r}_{n}(t) =[x_n(t),y_n(t)]$ where $n=1,...,N$, and possess orientations described by the unit vectors
$\bhu_n=(\cos\phi_n,\sin\phi_n)$. The self-propulsion velocity of each monomer is introduced 
via an effective driving force $\vec{F}_n=F\bhu_n$, acting along this orientation, see Fig.~\ref{fig:Sketch}(a). 
A single monomer experiencing overdamped dynamics with a translational friction coefficient $\gamma$ 
would then move with the constant self-propulsion speed $v_0 = F/\gamma$~\cite{BtHJPCM2015}.
We neglect any hydrodynamic interactions as in the traditional Rouse model \cite{Rouse_original}. 
The overdamped equations of motion for the positions $\vec{r}_n(t)$ of the $n$-th bead of the chain can then 
be written as

\begin{equation}
\gamma\frac{d\vec{r}_{n}}{dt}=-k \frac{\partial^2 \vec{r}_{n}}{\partial n^2}+ F\hat{\vec{u}}_n+\vec{\xi}_n(t)\,,
\label{eq:LangevinTrans}
\end{equation} 
whereby $\partial^2 \vec{r}_{n} / \partial n^2 = (\vec{r}_{n+1} + \vec{r}_{n-1} - 2\vec{r}_{n})$ corresponds to 
the spatial differential quotient of neighboring beads, using the constraints $\vec{r}_{0} = 0$ and $\vec{r}_{N+1} =0$.
Assuming the Stokes-Einstein relation applies not too far from equilibrium, 
the translational friction coefficient $\gamma$ is close to $\kbt/D$ 
 where $D$ is the short-time diffusion coefficient for a single bead and $\kbt$
 the thermal energy. The random forces $\vec{\xi}_n(t)$ are Gaussian distributed 
 with zero mean and variance 
$ \langle \vec{\xi} (t) \otimes \vec{\xi}(t^{\prime}) \rangle = 2\delta(t - t^{\prime}) \mathbbm{1} \left( \kbt \right)^2/ D $.
The orientation of the particle $\bhu_n$ is described by the rotational Langevin equation

\begin{equation}
\gamma_r \frac{d\hat{\vec{u}}_n}{dt}=\vec{\zeta}_n \times \hat{\vec{u}}_n(t)\,.   
\label{eq:LangevinRot}
\end{equation}
Here $\vec{\zeta}$  is a Gaussian-distributed torque with zero mean and variance
$ \langle \vec{\zeta} (t) \otimes \vec{\zeta}(t^{\prime}) \rangle = 2 \delta(t - t^{\prime}) \mathbbm{1} \left( \kbt \right)^2/ D_r$ 
and $\gamma_r$ is the rotational friction coefficient, which is close to $\kbt/D_r$. The rotational
diffusion coefficient $D_r$ is determined from the relation $D/D_r = 4b^2/3$.

In our model, $b = \sqrt{\kbt/k}$ characterizes a typical bead extension or bead distance
generated by the finite temperature, the so-called Kuhn length. In terms of this length, a typical elastic energy
contained in the harmonic spring is given by $kb^2$ such that we define the ratio between thermal and elastic energy
as the parameter 
\bea
\lambda = \frac{4 k b^2}{3 \kbt}\,.
\eea 
The strength of self-propulsion, on the other hand, is characterized by the P\'eclet number  
\bea
\mathrm{Pe}=\frac{v_0\, b}{D}\,.
\eea
 We do not consider explicit 
aligning interactions between neighboring beads.

In this paper we focus subsequently on three different situations, namely a  
 free  chain, a chain confined to an external
harmonic trap, and a chain dragged at one end. These situations are sketched and summarized in 
Fig.~\ref{fig:Sketch}.

A symmetric external harmonic trap potential is shown in Fig.~\ref{fig:Sketch}(b). It is  given by
\begin{equation}
U_{\text{ext}}(r) = \frac{1}{2}\kappa r^2,
\label{eq:Uext}
\end{equation}
and can be incorporated by adding the corresponding external force
$-\kappa \vec{r}_n$ to the right-hand-side of  Eq.~(\ref{eq:LangevinTrans}).
This introduces a further dimensionless parameter into the model, namely the ratio 
of the external and internal spring constants $\kappa/k$.

Lastly, a constant drag force $\vec{F}_d = -F_d \hat{\vec{e}}_x$ is applied to the first 
monomer of the chain, see Fig.~\ref{fig:Sketch}(c). The dragging force results
 in an anisotropic mean shape of the chain. The relative strength of the
 dragging can be described in terms of elasticity determined 
 by the dimensionless parameter $F_d/kb$.

\section{Computer simulations for a self-avoiding chain}
\label{sec:simu}
In the Brownian dynamics 
simulations, we model the active polymer as a sequence of $N$ coarse-grained spring beads in analogy to previous 
works considering short flexible rods~\cite{WanSoftMatter2013,SelfPropChain}.
 For simplicity, interactions between the active monomers are modeled by a smooth
repulsive WCA (Weeks-Chandler-Andersen) potential

\bea
U_{\text{WCA}}(r) = \begin{cases} 4 \epsilon \left[ \left(\frac{\sigma}{r}  \right)^{12} - \left( \frac{\sigma}{r}  \right)^{6}  \right] + \epsilon, &r \leq 2^{1/6}\sigma,\\0,  &r > 2^{1/6}\sigma. \end{cases}
\eea
Here  $\sigma$ denotes the diameter of a single bead and $\epsilon=\kbt$ is 
the interaction strength. These quantities represent the length and energy units, while times are conveniently measured
in units of the Brownian time $\tau = \sigma^{2}/D$. The self-avoidance of the chain is incorporated via the finite
length $\sigma$.

Springs between neighboring beads are introduced via a so-called FENE (finitely extensible nonlinear elastic) potential \cite{KremerMDPolymer1990}

\bea
U_{\text{FENE}}(r_{ij}) = - \frac{1}{2}K R_{0}^{2} \ln \left[ 1- \left(\frac{r_{ij}}{R_0} \right)^{2} \right],
\eea
with neighboring beads $i,j$ and their distance $r_{ij} = |\bbr_i - \bbr_j |$. The spring constant is
fixed to $K=27\epsilon/\sigma^{2}$ and the maximum allowed bond-length to $R_0 = 1.5\sigma$.

In our Brownian dynamics simulations, the same overdamped equations of motion [see  Eqs.~(\ref{eq:LangevinTrans})
and~(\ref{eq:LangevinRot})] were integrated with a small finite time step $\Delta t$, 
but now with self-avoidance and FENE-chains. The external harmonic potential and the dragged monomer were 
included using the same expressions as introduced in the previous chapter.

Statistics are gathered for up to 50 independent initial configurations along times of $t=10^4{\tau}$
after an equilibration period. The  
 time step used in our Brownian dynamics simulations is  $\Delta t = 10^{-4}\tau$. 
All our simulations are in two spatial dimensions.

\section{Results}
\label{sec:results}

\subsection{Free ideal chain}

First we focus on a free ideal chain which is long $(N\gg 1)$. In our analytical calculations, we 
consider the linear transformation
\begin{equation}
\vec{X}_p=\int_{1}^{N}dn\, \phi_{pn}\vec{r}_{n}(t)
\end{equation}
and choose the coefficients $\phi_{pn}$ such that the equation of motion for $\vec{X}_p$ has the form
\begin{equation}
\gamma_p\frac{d\vec{X}_p}{dt} 
=-k_p \vec{X}_p+ \vec{F}_p(t)\,. 
\label{transequ}
\end{equation} 
This leads to the relations\cite{Doi_Edwards_book}
\begin{align}
\phi_{pn}& =\frac{1}{N}\cos\left(\frac{p\pi (n-1)}{N-1}\right)\,,  \\
k_p& =k\frac{\gamma_p}{\gamma}\left(\frac{p\pi}{N}\right)^2 ,  \\
\vec{F}_p & =\frac{\gamma_p}{\gamma}\int_{1}^{N}dn\phi_{pn}\tilde{F}_n\,,
\label{transval}
\end{align}
using $\tilde{F}_n (t) = (F\hat{\vec{u}}_n+\vec{\xi}_n(t))\cdot \hat{\vec{e}}_{\alpha}$ and $\gamma_p = 2N\gamma$, with $p=1,2,\dots$
and $\alpha = x,y$. These equations can now be applied to calculate the mean square of the gyration radius as
\begin{align}
\big\langle{{R^2_G}_\alpha}(t)\big\rangle&=\Bigg\langle \frac{1}{N}\int_{1}^{N}dn {\left(\vec{r}_n-\vec{r}_\mathrm{cm}\right)_\alpha}^2\Bigg\rangle\nonumber\\
& =2\sum_{p=1}^{\infty}\big\langle X_p(t)X_p(t)\big\rangle\label{RG}\,.
\end{align}
Here the center of mass is given by $\vec{r}_\mathrm{cm}=\frac{1}{N} \sum_{n=1}^{N}{\vec{r}_n}$
and $\langle \cdots \rangle$ denotes a noise and time average. 
The end-to-end distance is given by 
\begin{align}
{R_E}_{\alpha}(t) = {r_N}_{\alpha}(t)-{r_1}_{\alpha}(t)=-4\sum_{\text{p: odd int}}X_p\label{end-to-end}
\end{align}
for each component, where the sum is over all odd values of $p$. 

\begin{figure}[thb]
\begin{center}
\includegraphics[width=1\columnwidth]{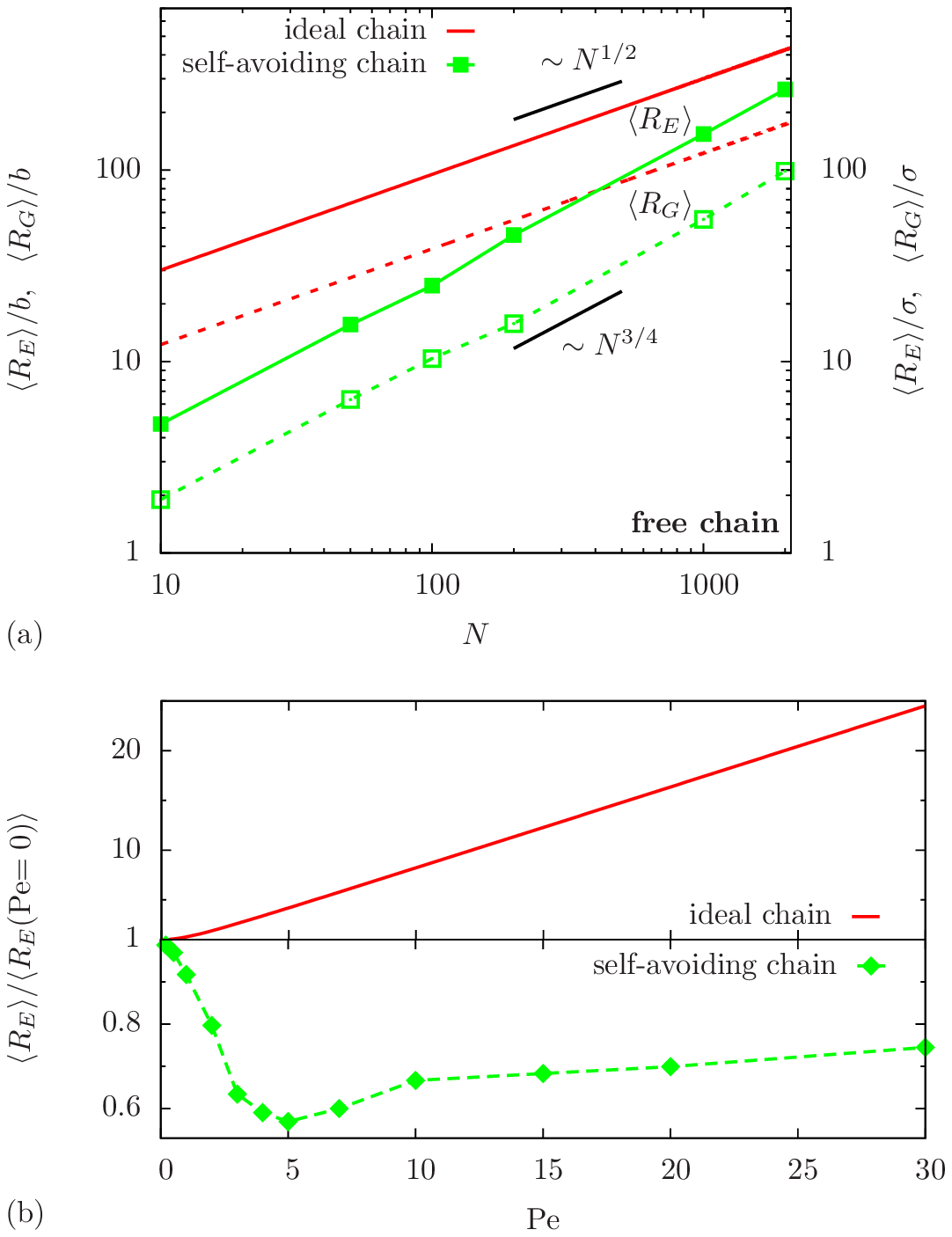}
\caption{\label{fig:FreeChain} (a) Reduced end-to-end distance $R_E$ (solid lines) and radius of gyration $R_G$ (dashed lines) for 
a free active polymer chain at $\text{Pe}=10$. (b) Relative change in $R_E$ induced by activity Pe for fixed number of active beads $N$.}
\end{center}
\end{figure}

For a long chain ($N\gg1$) of self-propelled particles we therefore find 

\begin{align}
\big\langle X_p(t)X_q(t)\big\rangle&=\left(\frac{{v_0}^2}{2}\frac{1}{k_p+D_r\gamma_p}\, \frac{\gamma_p\gamma}{k_p}+\frac{\kbt}{k_p}\right)\delta_{pq}\,, \nonumber \\
& \quad  p, q\ge 1\,,
\label{res1} \\
\big\langle {{R^2_G}}\big\rangle/b^2&= \frac{2}{9}\, \frac{N}{\lambda}\left(1+\frac{2}{3}{\mathrm{Pe}}^2\right)\,,
\label{rgfree} \\
\big\langle {{R^2_E}}\big\rangle/b^2 &=\frac{4}{3}\frac{N}{\lambda}\left(1+\frac{2}{3}{\mathrm{Pe}}^2\right)\,.
\label{refree}
\end{align}
Interestingly, the Flory exponent describing the 
basic scaling of spatial extension of the active polymer, see Eqs.~(\ref{rgfree}) and~(\ref{refree}), 
is the same as for a passive chain, i.e.\ we obtain for large $N$ both
\bea
\langle R_E^2 \rangle \sim N^{2\nu}
\eea 
and 
\bea
\langle R_G^2 \rangle \sim N^{2\nu}
\eea 
with $\nu=1/2$. This implies that for long chains without aligning interactions the effect of self-propulsion is not so dramatic
that the scaling is changed. We expect this to hold true even for short-ranged aligning interactions.
Long-ranged aligning interactions, however, could have an influence on the Flory scaling exponent, similar
to the scaling of polyelectrolyte chains with long-ranged Coulomb interactions~\cite{BarratJoanny}.

More precisely, the expressions~(\ref{rgfree}) and~(\ref{refree}) 
are identical to those for a passive polymer at a higher effective temperature
\begin{equation}
T_{\text{eff}}/T=1+\frac{2}{3}{\mathrm{Pe}}^2\,.
\end{equation}
The same mapping onto an effective temperature 
has been made for a single self-propelled 
monomer in a gravitational field~\cite{BocquetPRL2010} and has been tested for a single self-propelled bead
along a chain~\cite{Loi_SM1,Loi_SM2}. 

Finally, the long time diffusion coefficient for the chain is given by
\begin{align}
D_L=\frac{D}{N}\left(1+\frac{2}{3}{\mathrm{Pe}}^2\right)\,,
\end{align}
which is identical to the case of a single active colloid~\cite{tenHagen2009} but reduced by a factor $1/N$.

We now turn to effects of self-avoidance. In two spatial dimensions self-avoidance 
changes the Flory exponent from $1/2$ to $3/4$. Let us briefly recapitulate Flory's argument:
The free energy of a passive, self-avoiding chain in two dimensions is composed of two parts
\begin{eqnarray}
E \sim E_\text{ex} + E_\text{el}\,.
\label{eq:ScalingFree}
\end{eqnarray}
The first one, $E_\text{ex}\sim \sigma^2 N^2 /R^2$, arises due to excluded volume effects and
leads to chain swelling. The second term $E_\text{el}\sim k R^2 / N$ incorporates the elastic properties 
of the chain~\cite{FloryBook} and leads to chain shrinkage.
In equilibrium the total energy becomes minimal, leading to the well known Flory exponent $\nu=3/4$.

We confirm by simulations that 
the scaling exponent of the spatial extension, measured by $\langle R_E^2 \rangle$ or $\langle R_G^2 \rangle$,
is indeed $3/4$ unaffected by the activity, see Fig.~\ref{fig:FreeChain}(a)
~\footnote{The statistical error of the simulation data for an ideal chain is very small such that the
theoretical prediction and simulation are indistinguishable on the scale of the plots.}. 
This is in agreement with the experimental findings of an active granular chain~\cite{ExpGranularActiveChain1}, where for
chain lengths $N \leq 128$ the Flory exponent of a self-avoiding chain $\nu = 3/4$ has been found.

A striking difference generated by self-avoidance, however, is the P\'eclet number dependence of the chain extension.
As documented in Fig.~\ref{fig:FreeChain}(b), we find a non-monotonicity of the chain extension for increasing
 self-propulsion.
The chain first shrinks and then swells. The initial chain shrinkage can be attributed to self-propulsion 
of a particle encaged by its hard neighbors. Activity will on average drive the particle away from its neighbors which will
contribute to a zig-zag configuration of the chain implying a shrinkage. Increasing the activity further
will finally swell the chain since the larger driving force will extend the chain. The minimum in the chain size therefore
occurs when the self-propulsion force $F$ is getting comparable to the elastic force acting on a particle.
Interestingly, very recently, this effect has ben found 
in the reverse set-up of a passive chain in an active bath~\cite{KaiserPolymer,ValerianiPolymer} showing
that an active bath particle can be viewed as forming a joint unit with the polymer chain once 
it is colliding with the chain.

\subsection{Harmonically confined chain}

Now, we include a confining harmonic potential, see Eq.~(\ref{eq:Uext}). 
We  vary the prefactor $\kappa$ with respect to the spring
constants keeping the activity of the monomers fixed.

\begin{figure}[thb]
\begin{center}
\includegraphics[width=1\columnwidth]{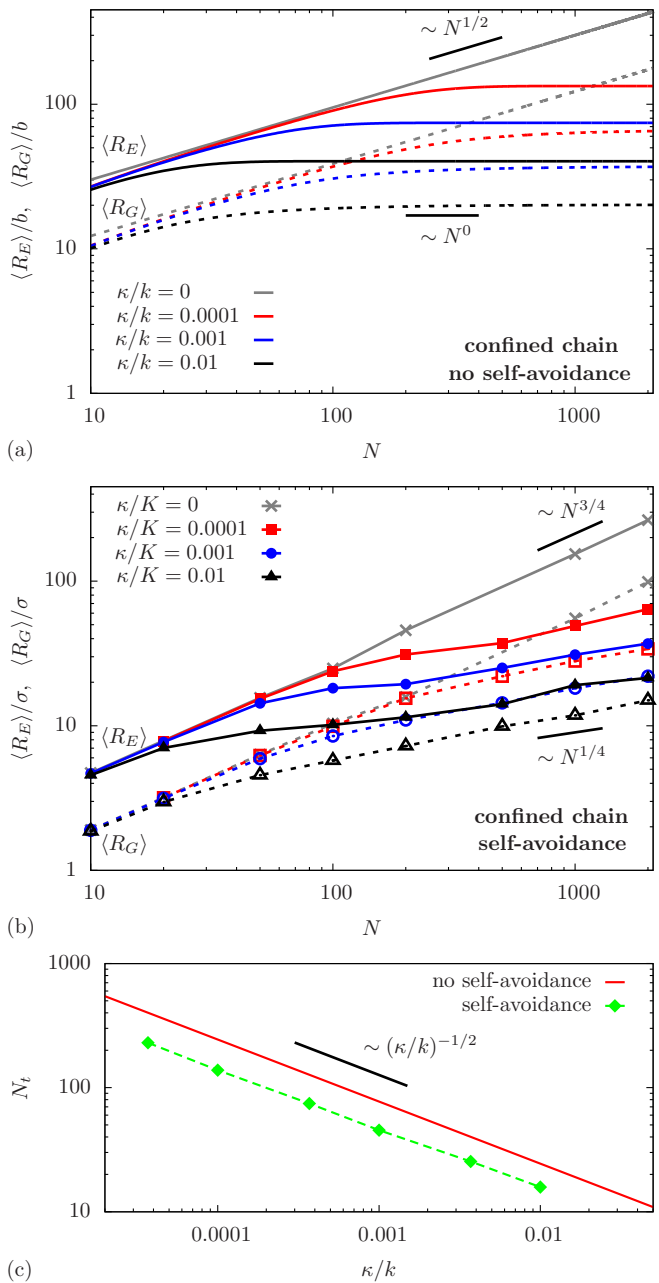}
\caption{\label{fig:HarmConf} Reduced end-to-end distance $R_E$ (solid lines) and radius of gyration $R_G$ 
(dashed lines) for a  harmonically confined polymer in the case of (a) no self-avoidance and (b) a self-avoiding chain 
for fixed activity $\text{Pe}=10$ of the monomers and varied confinement strengths $\kappa$.
(c) Threshold number of monomers $N_t$ as a function of $\kappa/k$.} 
\end{center}
\end{figure}

An analytical calculation for the ideal chain is still possible. However, the external 
potential leads to a shift in the values of $k_p$,
which are now given by
\begin{equation}
k_p= \frac{\gamma_p}{\gamma}\, \left[k\left(\frac{p\pi}{N}\right)^2 + \kappa \right] \,.
\end{equation}
Accordingly, the radius of gyration and the end-to-end distance are now given by
\begin{align}
\langle &{R^2_G}\rangle/b^2=\frac{4}{9 \lambda}\, \frac{k}{\kappa}\, {\mathrm{Pe}}^2
\Bigg[\sqrt{\frac{\kappa }{k}}\coth\left(N\sqrt{\frac{\kappa}{k}}\right)-\frac{1}{N}\nonumber\\
&-\frac{\lambda \kappa}{\lambda \kappa +k}\left(\sqrt{\frac{\lambda\kappa+k}{\lambda k}}\coth\left(N \sqrt{\frac{\lambda\kappa+k}{\lambda k}}\right)-\frac{1}{N}\right) \Bigg] \nonumber \\ 
& +\frac{2}{3\lambda}\frac{k}{\kappa}\left(\sqrt{\frac{\kappa}{k}}\coth\left(N\sqrt{\frac{\kappa }{k}}\right)-\frac{1}{N}\right)
\end{align}
and 
\begin{align}
\langle &{R^2_E}\rangle/b^2 =\frac{16}{9 \lambda}\, {\mathrm{Pe}}^2
\Bigg[-\sqrt{\frac{\lambda k}{\lambda \kappa +k}}\tanh\left(\frac{N}{2}\, \sqrt{\frac{\lambda \kappa +k}{\lambda k}}\right)\nonumber\\
& \quad +\sqrt{\frac{k}{\kappa}}\tanh\left(\frac{N}{2}\, \sqrt{\frac{\kappa}{k}}\right)\Bigg]
+\frac{8}{3\lambda}\sqrt{\frac{k}{\kappa}}\tanh\left(\frac{N}{2}\, \sqrt{\frac{\kappa }{k}}\right) \,,
\end{align}
respectively. 
In the limit of large $N$ we find the Flory exponent $\nu=0$ which implies that the chain is completely localized.
Again, the activity of the monomers only affects the 
prefactor but not the scaling behavior of the whole chain. 
As a function of monomer number $N$, a plateau in the spatial extension can be observed above a 
threshold chain length $N_t$ which scales with the strength of the applied external potential as
\bea
N_t = 2 \sqrt{\frac{k}{\kappa}} \left(1+\frac{2}{3}\text{Pe}^2 \right)^{-1}\,.
\label{eq:Nt_confinement}
\eea
Consequently, for large self-propulsion strengths, $\text{Pe}\gg1$, the threshold chain length decays as 
$N_t \sim \text{Pe}^{-2}$. 
Below this monomer number the well known Flory exponent $\nu=1/2$ is found, see Fig.~\ref{fig:HarmConf}(a), 
reproducing the trends for a passive polymer~\cite{James1991}.

For a self-avoiding passive polymer chain, Flory's argument, see Eq.~(\ref{eq:ScalingFree}), can be extended
by including an external harmonic potential to the total free energy as 
\begin{eqnarray}
E \sim E_\text{ex} + E_\text{el} + E_\text{ext}\,,
\label{eq:ScalingHarm}
\end{eqnarray}
where the scaling of $ E_\text{ext}$ can be obtained by integrating a constant sharp-kink monomer density 
profile inside a circle of radius $R$ as $E_\text{ext} \sim \kappa N R^2$.
The latter two contributions will enforce a compactification of the polymer
chain. 
For long chains $(N\gg1)$ the external energy $E_\text{ext}$ dominates the elastic energy which leads to 
a new exponent $\nu = 1/4$.  Below a threshold monomer number $N_t$ the elastic properties of the chain dominate such that
the Flory exponent is $\nu = 3/4$. Again this monomer number scales as $N_t \sim \sqrt{k/\kappa}$.

In case of a self-avoiding active polymer chain, we find the same Flory exponent $\nu = 1/4$ in the limit of large $N$,
see Fig.~\ref{fig:HarmConf}(b). As before, the scaling behavior is not drastically altered by the activity.
For varied confinement strenghts we can confirm the predicted scaling for the threshold monomer number $N_t$
for an ideal as well as a self-avoiding chain of active particles, see Fig.~\ref{fig:HarmConf}(c).

\subsection{Dragged chain}

\begin{figure}[thb]
\begin{center}
\includegraphics[width=1\columnwidth]{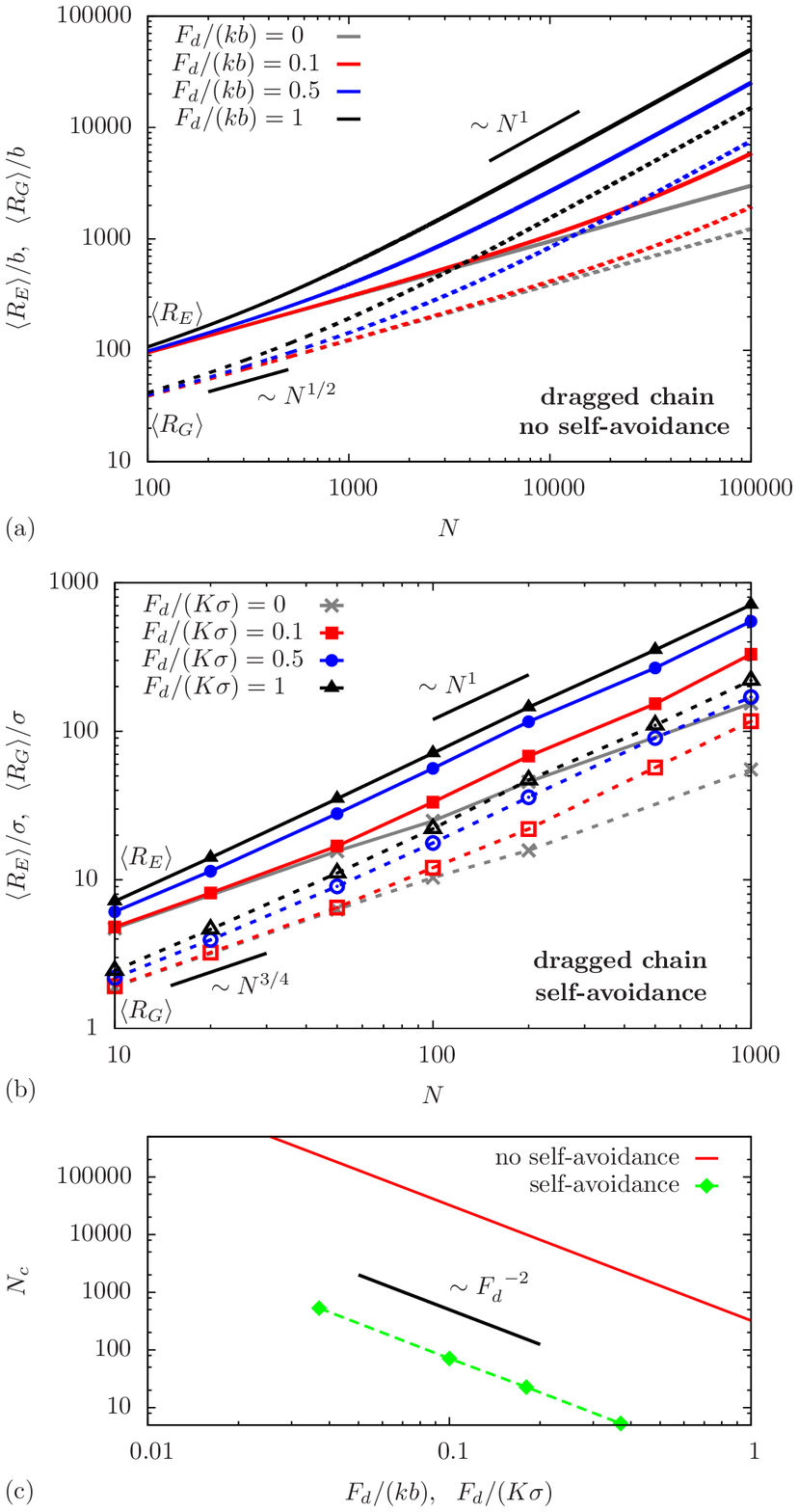}
\caption{\label{fig:DragChain} Reduced end-to-end distance $R_E$ (solid line) and radius of gyration 
$R_G$ (dashed line) for an active polymer which is dragged with a force $F_d$ 
in the case of (a) no self-avoidance and (b) a self-avoiding chain for a constant activity $\text{Pe}=10$.
(c) Crossover number of monomers $N_c$ for fixed activity as a function of reduced drag forces $F_d / (kb)$ respectively 
$F_d / (K\sigma)$.}
\end{center}
\end{figure}

We will now treat the situation of a constant drag force $F_d$ 
applied to the first monomer of the chain, see again Fig.~\ref{fig:Sketch}(c). 
The equations of motion are the same as in Eqs.~(\ref{eq:LangevinTrans}) and ~(\ref{eq:LangevinRot})
but now include the term $-F_d \hat{\vec{e}}_x$ of the right hand side of Eq.~(\ref{eq:LangevinTrans})
for $\vec{r}_1 (t)$.
The analytical solution of the Rouse model can be extended to an ideal dragged chain.
In detail, we obtain
\begin{align}
\big\langle X_p(t)X_q(t)\big\rangle&=\left(\frac{{v_0}^2}{2}\frac{1}{k_p+D_r\gamma_p}\ \frac{\gamma_p\gamma}{k_p}+\frac{\kbt}{k_p}\right)\delta_{pq} \nonumber \\  
& \quad +\frac{4\ {F_{d}}^2}{k_pk_q}\,, \qquad\qquad p, q\ge 1\,.
\end{align}
This implies that
\begin{align}
\big\langle R_G^2\big\rangle/b^2&=\frac{2}{9}\ \frac{N}{\lambda}\left(1+\frac{2}{3}{\mathrm{Pe}}^2\right)+\frac{1}{45}\left(\frac{F_{d}}{kb}\right)^2N^2 \label{eq:RGdrag}\,,\\
\big\langle R_E^2\big\rangle/b^2 &=\frac{4}{3}\frac{N}{\lambda}\left(1+\frac{2}{3}{\mathrm{Pe}}^2\right)
+\frac{1}{4}\left(\frac{F_{d}}{k b}\right)^2N^2\label{eq:REdrag}\,,
\end{align}
such that for long chains ($N\gg1$)
the Flory exponent is $\nu = 1$ in the dragged situation, see Fig.~\ref{fig:DragChain}(a). Intuitively this is expected as 
a strong  drag force stretches the chain along its pulling direction $\hat{\vec{e}}_x$. As becomes directly visible from the expressions 
(\ref{eq:RGdrag}) and (\ref{eq:REdrag}), there is a crossover in the scaling of the squared chain extension from $N$ to 
$N^2$. In fact, for small drag forces and small $N$ the Flory exponent is still that of an undragged chain ($\nu=1/2$).
For large self-propulsion ($\text{Pe}\gg1$), the crossover between the two regimes in $N$ occurs at a threshold of  
\bea
N_c \sim \left( \frac{\text{Pe}\,kb}{F_d}  \right)^2 \,.  
\label{eq:Nt_drag}
\eea
Our computer simulations for a self-avoiding chain confirm the same qualitative behavior.
Now there is a crossover from the Flory exponent $\nu=3/4$ of an ideal self-avoiding
chain to the stretched case $\nu=1$, see Fig.~\ref{fig:DragChain}(b), whereby the crossover number of monomers $N_c$
decreases as given in Eq.~(\ref{eq:Nt_drag}), see Fig.~\ref{fig:DragChain}(c). Again, the activity of the chain does not affect the
scaling exponent which stays to be  $\nu=1$ as known from a passive chain~\cite{Pincus1977,Sakaue_Dragging}. 

\begin{figure}[thb]
\begin{center}
\includegraphics[width=1\columnwidth]{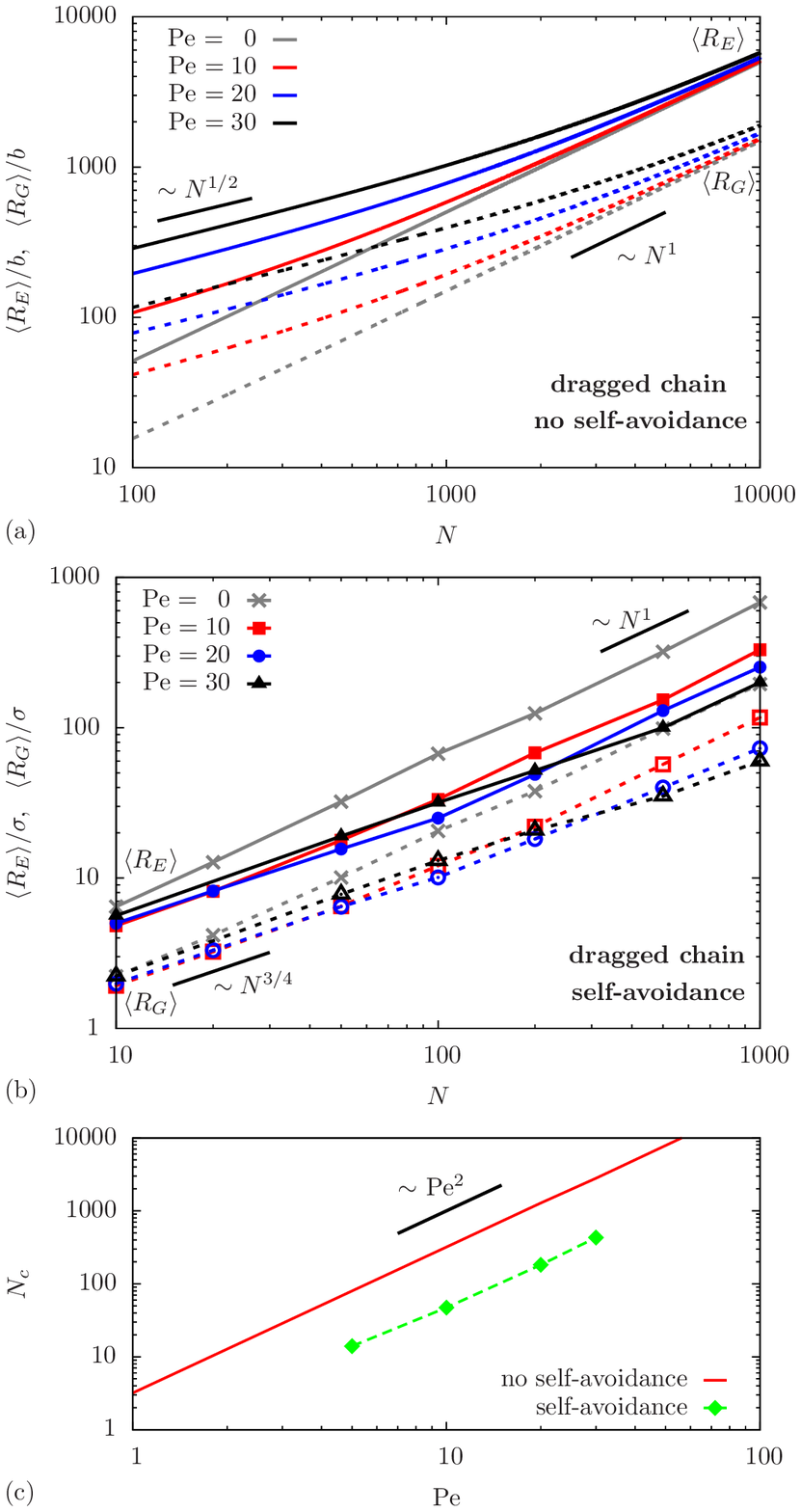}
\caption{\label{fig:FixedDrag} Reduced end-to-end distance $R_E$ (solid line) and radius of gyration 
$R_G$ (dashed line) for a chain of active particles which is dragged with a fixed force $F_d$ 
in the case of: (a) $F_d/(kb) = 1$ with no self-avoidance and (b) $F_d/(K\sigma)=1$ for a self-avoiding 
chain for varied activity Pe. (c) Crossover number of monomers $N_c$ for a constant drag as a function of activity Pe.}
\end{center}
\end{figure}

Finally, in Fig.~\ref{fig:FixedDrag}, we study the P\'eclet number dependence of the extension of a dragged chain.
Indeed the threshold chain length for the crossover $N_c$ scales as predicted in Eq.~(\ref{eq:Nt_drag}) and
shifts for higher self-propulsion to higher $N$, see Fig.~\ref{fig:FixedDrag}(c).

\section{Discussion and Conclusions}
\label{sec:conc}

We have generalized the traditional Rouse model of polymer dynamics to the situation of an active polymer consisting of
self-propelled monomers. By analytically solving the Rouse model and numerical simulations of a self-avoiding polymer
we have shown that the well known Flory scaling exponents are still valid in the presence of activity.
For an ideal chain of active particles, the activity only affects the prefactor and can be considered as an effective temperature
when mapped onto a corresponding passive polymer. In the case of a self-avoiding chain, the spatial extension reveals a 
non-monotonicity as a function of activity which is absent for an ideal chain.
The cases of harmonically confined and dragged active polymer chains reveal a crossover in scaling for the chain swelling but again activity 
does affect the Flory exponent for long chains.

Chains of active particles can be prepared in nature  by chaining artificial colloids. This can be achieved 
 using the lock-and-key technique~\cite{Pine_Nature_2010} or dipolar or patchy colloids that prefer chaining,
for some recent realizations, see Refs.~\cite{polloidal_chains,Herminghaus_SM14}. In principle, one can use active colloids
as entities to obtain a one-dimensional chain of active particles. This has not yet been realized but is at least conceivable.
The main advantage of the colloidal realization is that the motion of the chain can directly be observed in real-space
such that the statistical average needed for the mean size is obvious. Moreover the activity 
can be controlled from outside such that
a direct comparison with a passive thermal chain is possible, which enables a test of the effective temperature scaling.
Though our model is designed for Brownian motion in a solvent, we expect that
the key trends are the same as for driven granulates. 
Therefore, it would be interesting to perform more granulate experiments in confinement and under drag to test our further predictions.
A harmonic confining potential can easily be realized by shaking the granular not on the plane but within a paraboloid.
Finally, dragging the first monomer can be realized by charging one end-monomer using electrofriction~\cite{Whitesides_SM09}
and placing the whole granulate chain into a homogeneous electric field along the vibrating substrate.

Future research should consider three spatial dimensions where the description of the orientational dependence 
of Brownian motion is more complicated \cite{Wittkowski2012}, 
but similar results are expected. While the scaling should still correspond to that of a passive chain,
the prefactor will be affected by the activity. In particular, in three spatial dimensions the activity is less important,
which is already visible for the effective temperature mapping. For active spheres in three dimensions 
$T_{\text{eff}}/T = 1 + \frac{2}{9}\text{Pe}^2$ holds with a prefactor 2/9, instead of 
$T_{\text{eff}}/T = 1 + \frac{2}{3}\text{Pe}^2$ in two dimensions where the prefactor 2/3 is stronger~\cite{BtH2011}.
Another line of future research concerns a chain with both active and passive monomers. Here, again, we expect the same scaling exponent but a prefactor in the effective temperature mapping which is smaller than in the case considered here, where any monomer is active.
Moreover, the
influence of hydrodynamic interactions needs to be considered in more detail following the work of
 Ref.~\cite{Jayaraman_2012} which applies to short and stiff filaments. If the solvent is viscoelastic, 
new effects come into play, see e.g. Refs.~\cite{ElasticConf,Fischer_NC2014}, which also need to be explored in the future.

\acknowledgments
We thank Andreas M. Menzel for helpful discussions. This work was financially supported by the ERC Advanced Grant INTERCOCOS
(Grant No. 267499) and by the SPP 1726 of the DFG.

\bibliography{ref}

\end{document}